\documentclass[a4paper]{jpconf}
\usepackage{graphicx}
\begin{document}
\def\nn{\nonumber \\}
\def\p{\partial}
\def\t{\tilde}
\def\h{{1\over 2}}
\def\be{\begin{equation}}
\def\bea{\begin{eqnarray}}
\def\ee{\end{equation}}
\def\eea{\end{eqnarray}}
\def\g{\Gamma}
\def\hint{\hat{\int}}
\def\w{{\tilde\omega}}
\def\sqi{{1\over \sqrt{2}}}
\def\norm{{1\over 2\pi i}}
\def\r{\rightarrow}
\def\b{\bigskip}

\title{Black holes and holography${}^{*}$}\footnote{Expanded version of proceedings for COSGRAV12, Kolkata, Feb 2012.}
\author{Samir D. Mathur}

\address{Department of Physics, The Ohio State University, 191 W. Woodruff Ave, Columbus, OH 43210, USA}

\ead{mathur.16@osu.edu}

\begin{abstract}
The idea of holography in gravity arose from the fact that the entropy of black holes is given by their surface area. The holography encountered in gauge/gravity duality has no such relation however; the boundary surface can be placed at an arbitrary location in AdS space and its area does not give the entropy of the bulk. The essential issues are also different between the two cases: in black holes we get Hawking radiation from the `holographic surface' which leads to the information issue, while in gauge/gravity duality there is no such radiation from the boundary surface.
 To resolve the information paradox we need 
to show that there are real degrees of freedom at the horizon of the hole; this is achieved by the fuzzball construction. While the fuzzball has no interior to the horizon, we argue that an auxiliary spacetime can be constructed to continue the collective dynamics of fuzzball for times of order the crossing time; this is an analogue of `complementarity'.
\end{abstract}

\section{Introduction}

The idea of holography arose in gravitational physics from the expression for the  entropy of black holes \cite{bek}
\be
S_{bek}={A\over 4G}
\label{one}
\ee
Since the entropy  is given by the surface area measured in units of $l_p^2$, it appeared plausible that there is `one bit of data per planck area of the horizon'. Because the degrees of freedom are given by the bounding area rather than the `volume' of the hole, we use the term `holographic' to characterize the gravitational physics of the hole. 

\begin{figure}[htbt]
\hskip .5 in\includegraphics[scale=.35]{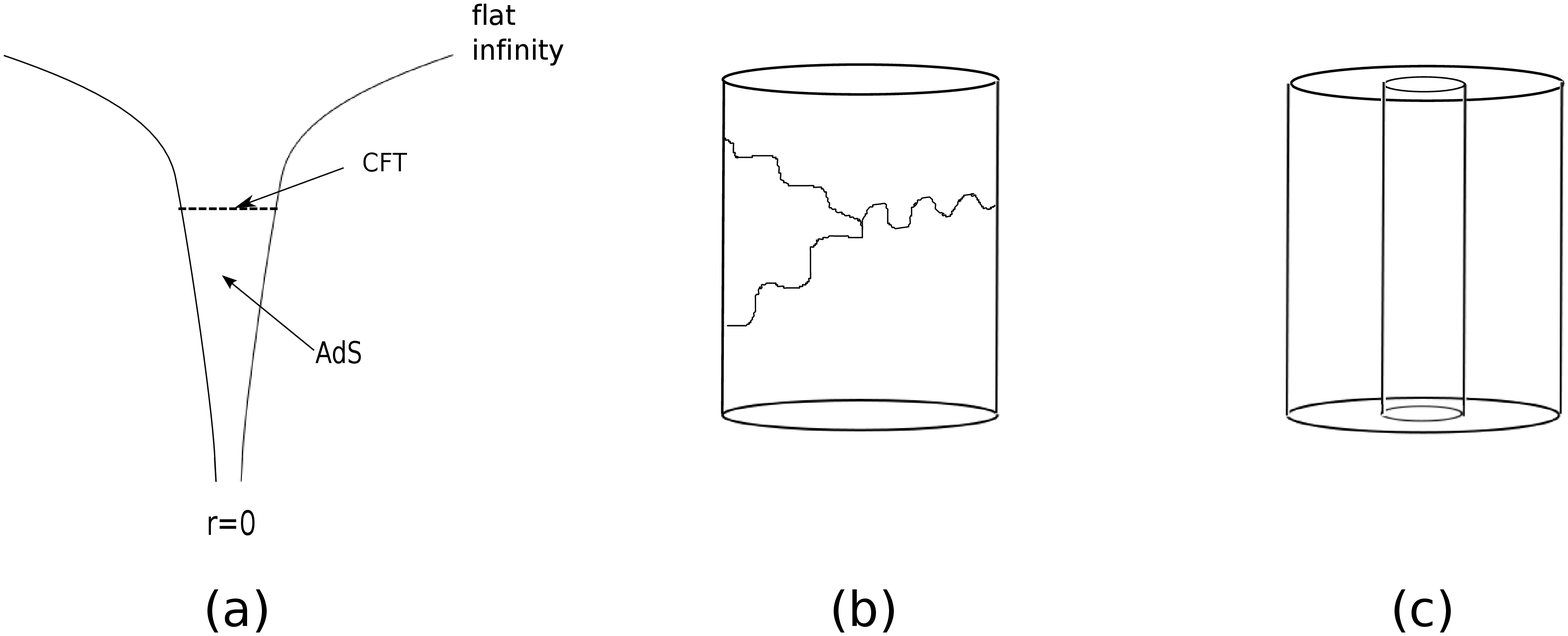}
%
%
\caption{(a) Branes create a geometry that is AdS in the `near' region; the dual CFT lives on a boundary placed anywhere in the AdS region, and describes gravity in all the region below it. (b) The singularity at $r=0$ can be avoided by moving to global AdS, where a 3-point function is computed by a simple path integral with no singularities. (c)  If we have enough energy in global AdS, we make a black hole, and then we face the difficulties of Hawking's argument again. (The vertical direction is time, and the surface of the inner cylinder is the black hole horizon.)}
\label{fa1}       
\end{figure}

In recent years the term `holography' has been applied to the idea of gauge/gravity duality in AdS spacetimes \cite{maldacena}. This idea is depicted in fig.\ref{fa1}(a). The dashed line is an arbitrary boundary surface ${\cal S}$. A  field theory on ${\cal S}$ describes the gravitational physics of the entire region below ${\cal S}$. Again, it appears that physics is `holographic'.

Since the idea of gauge/gravity duality arose from studies of black holes in string theory, it is often assumed that these two uses of the term `holographic' are the same. Stretching this further, one may think that the idea of gauge/gravity duality would somehow explain the mysteries associated with the entropy (\ref{one}) and the related problem of the information paradox \cite{hawking}. But as we now note, there are several differences between the above two notions of holography:

\b

(i) For the black hole, the location of the boundary surface (the horizon)  is  fixed at $r=2M$. In gauge/gravity duality, the boundary surface can be placed anywhere in the AdS region.

(ii) For the black hole, the area of the boundary surface gives the entropy (\ref{one}) of the interior. But in gauge/gravity duality there is no such relation between the area of the boundary and the entropy contained inside. such relation. The entropy of the interior spacetime depends on how much energy $E$ we put in it. In particular if we take $E=0$, we get empty AdS with entropy $S=0$.

\b

As we probe these differences more deeply, we will uncover important aspects of  gravitational degrees of freedom and of the structure of spacetime itself.

\section{Lessons from the information paradox}

\begin{figure}[htbp]
\begin{center}
\includegraphics[scale=.18]{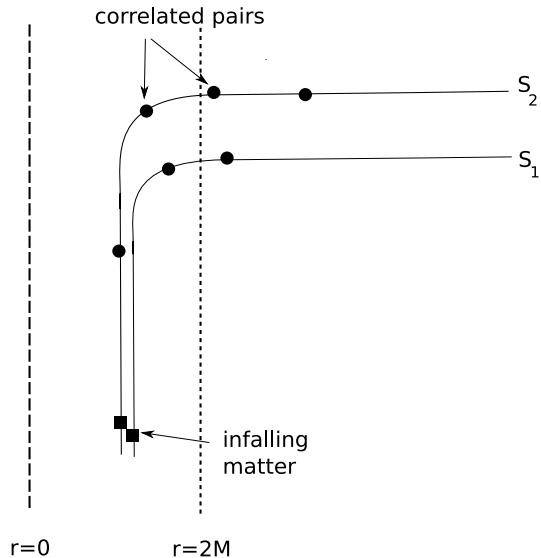}
\caption{{Eddington-Finkelstein coordinates for the Schwarzschild hole. Spacelike slices are $t=const$ outside the horizon and $r=const$ inside.  Curvature length scale for a solar mass black hole is $\sim 3 ~km$ all over the region of evolution covered by the slices $S_i$.}}
\label{ftwo}
\end{center}
\end{figure}

A fundamental problem in black hole physics is the information paradox \cite{hawking}.  In a geometry with horizon, there is no time-independent foliation of spacetime. As shown in fig.\ref{ftwo}, spacelike slices are $t=constant$ outside the horizon, but $r=constant$ inside.  `Later' slices are obtained by `stretching' the $r=constant$ part of the slice; this stretching creates particle pairs, with one member being inside the horizon and one outside. The important aspect of this particle creation process is that the two members of the pair are in an {\it entangled} state, which we may write schematically as
\be
|\psi\rangle_{pair}={1\over \sqrt{2}}\left ( |0\rangle_{in}|0\rangle_{out}+|1\rangle_{in}|1\rangle_{out}\right )
\label{three}
\ee
where $ |0\rangle,  |1\rangle$ represent occupation numbers 0, 1 respectively for a given particle mode. Thus each step of the pair creation process generates an entanglement entropy $S_{ent}=\ln 2$, and after $N$ steps of particle creation the entanglement is
\be
S_{ent}=N\ln 2
\label{two}
\ee

This pair creation continues until the hole reaches planck size, at which point the emitted radiation is heavily entangled with the quanta in the hole. If the hole evaporates completely, the radiation cannot be attributed {\it any} quantum state -- the radiation is entangled, but there is nothing that it is entangled {\it with}. This possibility is termed `information loss' or `loss of unitarity'. If we are left with a remnant, then this remnant needs to have at least $2^N$ internal states to permit the entanglement (\ref{two}). Since $N$ is unbounded, our theory must permit arbitrarily high degeneracy in a bounded volume with bounded energy budget; something that is hard to achieve with normal quantum theories.

Many relativists had reconciled themselves to admitting remnants of some sort, perhaps with baby Universes opening up inside the remnant where states could be hidden \cite{baby}. But such remnants have not been found in string theory. How do we avoid being forced to `information loss', which would imply a breakdown of string theory?

Some string theorists have been seriously worried about this problem. But many others assumed that Hawking's argument was somehow {\it flawed}. Among the latter, the most common belief was the following. Hawking computed the pair creation at leading order, but there can always be small quantum gravity corrections to the wavefunction (\ref{three})
\be
|\psi\rangle_{pair}={1\over \sqrt{2}}\left ( |0\rangle_{in}|0\rangle_{out}+|1\rangle_{in}|1\rangle_{out}\right )+\epsilon {1\over \sqrt{2}}\left ( |0\rangle_{in}|0\rangle_{out}-|1\rangle_{in}|1\rangle_{out}\right )
\ee
where we have added a small amount of an orthogonal state for the pair. The correction $\epsilon$ for each pair must be small since the horizon geometry is smooth. But the number of emitted quanta is large ($\sim (M/m_p)^2$), and one might hope that the net effect of the small corrections can cumulate in such a way that the overall state of the radiation would be un-entangled with the hole. In that case, of course, there would be no real information paradox to worry about. 

But in \cite{cern} it was shown that this hope is false; the change in entanglement is bounded as
\be
{\delta S_{ent}\over S_{ent}}<2\epsilon
\label{four}
\ee
This inequality is the essential reason why the Hawking argument has proved so robust over the years -- no small corrections can save the situation.

To summarize, it can be rigorously argued that we must choose between one of the following: (i) we have information loss or remnants (ii) we find a way to get order {\it unity} corrections to low energy physics at the horizon.

\section{Gauge/gravity duality and the information paradox}\label{sec2}

With the above knowledge of the information problem, let us return to our analysis of holography. A common error is to argue the following: ``Many computations support the idea that gravity is dual to a gauge theory. Since the latter is unitary, there cannot be information loss in black holes. Thus we have solved the information problem".

As we will now see, this argument is completely incorrect, and arises from ignoring the power of the information paradox, encoded in (\ref{four}). We can approach gauge/gravity duality from two sides:

(i) We know string theory at low energy gives gravity. Low energy gravity ampltiudes can be reproduced by the gauge theory. But if we put together enough energy to make an AdS-Schwarzschild black hole, then the Hawking argument tells us that we will get information loss. Thus unless we find some way to bypass the Hawking argument, {\it gauge/gravity duality would fail at the same place where all other quantum gravity approaches fail: at the threshold of black hole formation}. 

(ii) We can {\it define} the gravity theory as the dual of the gauge theory. In this case we cannot lose unitarity. {\it But now we cannot assume that the dual gravity theory has black holes.} Low energy amplitudes in the gauge theory agree with gravity. But if we take a large energy excitation in the gauge theory, then the natural timescale for dispersion of this energy is the {\it crossing time} of the black hole, not the much longer Hawking evaporation time. If the energy disperses in order crossing time, we have no black hole in the theory. 

\b

As a concrete illustration of the above points, we can look at the simplest manifestation of a gauge/gravity type correspondence: the 1-d matrix model which is dual to 1+1 dimensional gravity \cite{matrixmodel}. The low energy gravity theory is 1+1 dimensional dilaton gravity, in which we can make a black hole by throwing  matter towards $r=0$ \cite{cghs}. With this black hole  we find information loss or remnants, depending on how we complete the theory at the planck scale.
The 1-d matrix model, on the other hand, gives a theory that must be unitary by construction. Low energy amplitudes computed with the matrix model indeed reproduce the amplitudes of  1+1 dimensional dilaton gravity. But if we try to make a black hole with the matrix model then we {\it fail}: the energy of a collapsing shell bounces off the origin and returns in a time of order {\it crossing time}. This difference in behavior between the matrix model and dilaton gravity is  caused by higher order quantum corrections, which are small at low energies, but grow large enough at the black hole threshold to prevent black hole formation altogether \cite{matrixbh}. 

In short, gauge/gravity duality has no direct bearing on the information problem. As we noted in the introduction, the `holography' in gauge/gravity duality is not the same as the `holography' that arises for black holes from  the entropy formula (\ref{one}). The holography of gauge/gravity duality can be verified for low energy correlators, but above the threshold of black hole formation the duality cannot answer any quantum gravity questions until we understand the nature of gravitational states in this domain. The traditional approach of writing down the AdS-Schwarzschild metric for the black hole allows us to define thermal averages in the dual gauge theory, but cannot tell us anything about how the information paradox is to be resolved. For the latter question, we need to understand black hole microstates themselves.

\section{Resolving the information paradox -- fuzzballs}\label{sec5}

In string theory, we have to make black hole states from the objects present in the theory -- strings and branes. It turns out that the size of brane bound states grows with the coupling and with the number of branes in the bound state -- in such a way that the size of the state is always order horizon size \cite{emission}. Thus we do not get a traditional horizon with vacuum in its vicinity, as was assumed in the Hawking computation; instead we get a horizon sized `fuzzball'.  The emission of low energy modes can  therefore be modified by order unity, as required to solve the information paradox. 

Further progress along these lines is obtained by taking specific states of the hole and constructing their gravity description. In each case we see that no horizon forms. It is interesting to see these constructions in the context of the no-hair conjectures that suggested that the traditional black hole geometry was {\it unique}. In string theory we have extra compact dimensions, and a set of sources (branes). These objects are all crucial to the structure of microstates. A compact circle `pinches off' before reaching the horizon, creating a smooth end to the spacetime geometry. This pinch-off generates a set of Kaluza-Klein monopoles and antimonopoles just outside the place where the horizon would have formed, and fluxes corresponding to brane charges wrap the topological cycles produced by this monopole structure. The simplest black hole is the extremal 2-charge hole made from D1 and D5 branes. For this hole we find the following:

 (i) The number of extremal bound states of the D1D5 brane system can be counted by abstract topological methods, and give a microscopic entropy $S_{micro}=4\pi\sqrt{n_1n_5}$ \cite{sen}.
 
 (ii) If we assume a spherically symmetric ansatz and a trivial factorization of the compact directions, then the low energy supergravity action gives an extremal black hole with horizon, with a Bekenstein-Wald entropy $S_{bek}=4\pi\sqrt{n_1n_5}$ \cite{dabholkar}.
 
 (iii) The actual microstates of the D1D5 system can be constructed.  It is found that they are not spherically symmetric and the compact directions are locally nontrivially fibered, though the net monopole charge of these fibrations vanishes. The solutions have no horizon and no singularity \cite{lm4,fuzzball2}.
 
 (iv) The phase space of these horizonless gravitational solutions can be quantized, and yields the entropy $S=4\pi\sqrt{n_1n_5}$ \cite{rychkov}.
 
 (v) Though there is no horizon for any microstate, the  region where the typical microstates  exhibits their nontrivial structure has a boundary whose area $A$ satisfies $A/G\sim \sqrt{n_1n_5}\sim S_{bek}$ \cite{lm5}.

\b

Work on more complicated extremal holes \cite{sv}  has yielded a similar picture \cite{fuzzball3}, though all microstates have not been constructed yet. Some families of nonextremal microstates have been constructed as well \cite{ross}, and again they have no horizon or singularity. But they do have {\it ergoregions}, and the rate of ergoregion emission \cite{myers} agrees exactly with the Hawking radiation rate expected for these microstates \cite{cm1}. But this time the radiation process does not lead to information loss; the radiation is similar to that from a normal warm body.

\section{The fate of a collapsing shell}
 
 Let us pause for a moment to see what the above discussion says about holography. We again find a difference between the case of the black hole and the case of gauge/gravity duality:
 
 \b
 
 (a) For the black hole, we have found that microstates do not have a `traditional' horizon. Here we use the term `traditional horizon' to denote the kind of structure that had been historically assumed for black holes: a boundary from inside which light rays cannot escape, with smooth spacetime  in an open set around this boundary.  Instead, we have found a {\it fuzzball};  the information of the microstate is encoded in the detailed structure of the microstate at the location where the horizon would have been. In short, there are {\it real} degrees of freedom at the surface which is used in the holographic expression (\ref{one}).

 (ii) In the case of gauge/gravity duality, there are no degrees of freedom apparent at the location of the boundary used for holography. This fact is related to the observation that the holographic boundary in this case can be moved to an arbitrary location in the AdS region. 
 
 \b
 
We now turn to addressing a  common question with black holes: what happens to a shell that is collapsing to make a black hole?

Consider a shell of mass $M$ that is collapsing through its horizon radius $R\sim GM$. In ordinary 3+1 dimensional gravity  the wavefunction of the shell moves in the expected way to smaller $r$, creating the traditional black hole geometry. But in string theory the $e^{S_{bek}}$ fuzzball states of the hole give alternate wavefunctions with the same quantum numbers as the shell. There is a small amplitude for the wavefunction of the shell to tunnel into one of these microstate wavefunctions. We may estimate this amplitude  as ${\cal A}\sim e^{-S_{gravity}}$ where  $S_{gravity}={1\over 16\pi G}\int {\cal R} \sqrt{-g} d^4 x$ and we use  $\sim GM$ for all length scales. This gives
\be
S_{gravity}\sim {1\over G}\int {\cal R}\sqrt{-g} \, d^4 x \sim {1\over G} {1\over (GM)^2}(GM)^4\sim GM^2\sim \Big ( {M\over m_{p}}\Big ) ^2
\ee
Thus ${\cal A}\sim e^{-(M/m_p)^2}$ is indeed tiny. But we must multiply the tunneling probability by the {\it number} of states that we can tunnel to, and this is given by ${\cal N}\sim e^{S_{bek}}$ where
\be
S_{bek}={A\over 4G}\sim {(GM)^2\over G}\sim \Big ( {M\over m_{p}}\Big ) ^2
\ee

 Thus the smallness of the tunneling amplitude is offset by the remarkably large degeneracy of states that the black hole has \cite{tunnel}. The wavefunction of the shell tunnels into these fuzzball states in a time much shorter than the evaporation time of the hole \cite{rate}, and then the fuzzballs states radiate energy much like any other normal body. In short, the semiclassical approximation that leads to the standard black hole geometry  gets invalidated by the large measure of phase space over which the wavefunction of the shell can spread. 

\section{The failure of the `good slicing' argument}

\begin{figure}[htbt]
\hskip .5 in\includegraphics[scale=.35]{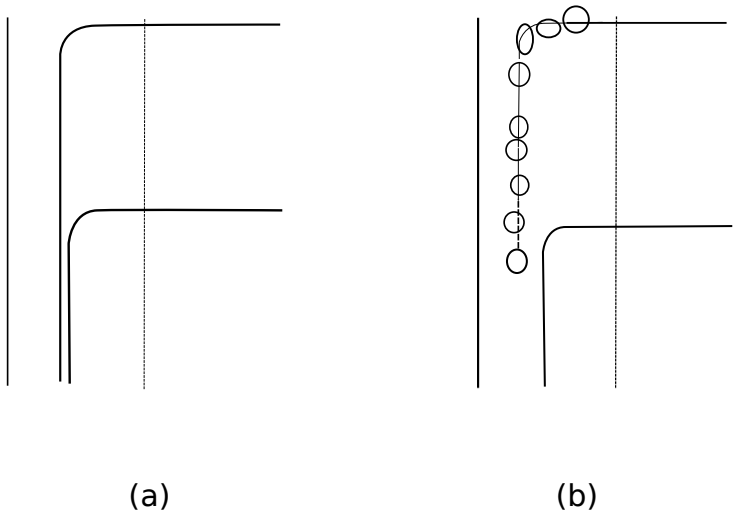}
%
%
\caption{(a) The stretching of `good slices' in the traditional black hole geometry leads to pair creation by the Hawking process and the consequent information problem. (b) If there are $Exp[S]$ fuzzball solutions, the wavefunction giving semiclassical geometry on the initial slice spreads over this vast phase space of solutions after some evolution, and we no longer get the traditional pair creation with growing entanglement.}
\label{fa13}       
\end{figure}

We can rephrase the above discussion in the following way. The information problem arose due to the `good slicing' of the black hole, in which we keep the slice smooth while stretching it more and more (fig.\ref{ftwo}). This stretching creates the entangled pairs that lead to Hawking's paradox. But with the above estimate of tunneling rates, we have a  different situation. On an early time slice we can indeed arrange the wavefunction so that it describes a semiclassical spacetime containing a collapsing shell. But the wavefunction of a later slice has to be obtained by evolution (using the Hamiltonian constraint) from the wavefunction on the initial slice. If we had a simple theory of quantum gravity, like canonically quantized general relativity, then evolution of the earlier slice would indeed give the `stretched' slice. But in string theory the situation is different. We have an enormous space of alternative solution to the gravity theory, where for example the compact directions can `pinch off' to make monopole pairs. We can take the wavefunction on the initial slice to be peaked around the smooth gently curved manifold, but the evolution will force this initial wavefunction to spread over the space of fuzzball solutions. Thus instead of getting the `stretched' slice, we get a linear combination of fuzzballs, which then radiate from their surface just like any other warm body. We depict this spreading of the wavefunction in fig.\ref{fa13}. 

We can also use this picture to address some related questions. Marolf \cite{marolf} has discussed the information problem in the following language. Suppose we assume that the gravity theory is holographically dual to a boundary field theory. Then the boundary theory contains all the information in the bulk, and we can connect the bulk state at early times to the bulk state at late times by just evolving unitarily in the boundary theory. Thus it seems impossible to have information loss in any theory with a gravity dual. 

While this argument helps to frame the information problem in the context of gauge/gravity duality, note that we cannot use it to argue away the  paradox itself. The reasons are the same as those we discussed in section \ref{sec2}. If we start with some simple theory of gravity, then we do not know that it will have a holographic dual  at the energies where black holes will form. If we start with the  field theory and {\it define} bulk gravity as its dual, then we do not know that we will get black holes.

But Marolf's argument {\it can} be used to sharpen the puzzles arising from the paradox, as was done recently in a paper by Heemskerk, Marolf and Polchinski \cite{hmp}. These authors asked if the dual field theory captured the state of a `Schrodinger's cat' that was behind the horizon. At early times when there was no black hole, the dual field theory presumably did capture all details of the bulk, and therefore captured the state of the cat. Evolution in the boundary is a known unitary evolution, so if the cat then evolved to be at a point inside the horizon, then its state would also be captured by the boundary theory. What appears puzzling though is that the radiation emitted from the hole should {\it also} capture the information of the cat, and one can draw a Cauchy slice that captures both this radiation and the cat behind the horizon. Since information cannot be duplicated, one expects a kind of `complementarity' \cite{complementarity} which somehow does not create a contradiction between the presence of these two copies. This argument, of course, is just a restatement of the usual information problem of black holes; it appears sharper in the context of the dual field theory since this field theory is manifestly unitary and does not duplicate information.

With the help of fig.\ref{fa13} we can see how the puzzle raised in \cite{hmp} is resolved with fuzzballs. Suppose the initial slice contains the collapsing shell and the cat, in their usual semiclassical wavefunctions. But the slice that captures a significant part of the radiation involves a lot of stretching, and this evolution spreads the wavefunction away from semiclassical slices to a linear combination of fuzzballs. The data of the cat is encoded in these fuzzballs, and is then carried away in the radiation from the fuzzballs. This latter radiation is just like radiation from any warm body, as shown by the explicit ergoregion emission computations mentioned in section \ref{sec5}.

\section{Dynamics of fuzzballs}

\begin{figure}[htbp]
\begin{center}
\includegraphics[scale=.15]{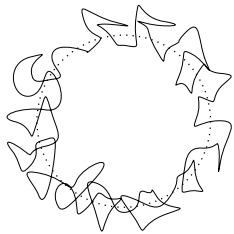}
\caption{Schematic description of a microstate solution of Einstein's equations. There are `local ergoregions' with rapidly changing direction of frame dragging near the horizon. The geometry closes off without having an interior horizon or singularity due to its peculiar topological structure.}
\label{ftwoq}
\end{center}
\end{figure}

We have seen that in string theory the microstates of black holes do not have a horizon; the state ends in a quantum mess of string theory sources just outside the place where the horizon would have been. For simple microstates we have seen that Hawking radiation emerges from the ergoregion in the geometry; the rate of this radiation was exactly what was expected from this microstate. But there is no information loss, since the other member of the produced pair does not fall through a horizon; instead it sits in the ergoregion, where it influences the production of future quanta. This is exactly the behavior required for an information carrying evaporation process. Equivalently, we can probe the  fuzzball with low energy quanta and discover its state. For example, quanta sent onto the ergoregion will lead to an enhanced stimulated emission from the ergoregion, so we can find the ergoregion spectrum. In short, we have a situation similar to that envisaged in the `membrane paradigm' \cite{membranebook}, but the degrees of freedom at the horizon are {\it real} \cite{membrane}. We depict this in fig.\ref{ftwoq}.

If we cannot  go `through the horizon', then is there any significance to the traditional black hole geometry where we can continue past the horizon and fall into a singularity? In \cite{plumberg} it was argued that there is a way to understand the extended Schwarzschild black hole geometry as an  auxiliary spacetime which can be used to reproduce an approximation to appropriate correlators in the fuzzball solution. We describe this conjecture below. But first we need to describe a few background steps, for which we briefly summarize some earlier results.

\section{Entangling gravity states}

  \begin{figure}[htbp]
\begin{center}
\includegraphics[scale=.65]{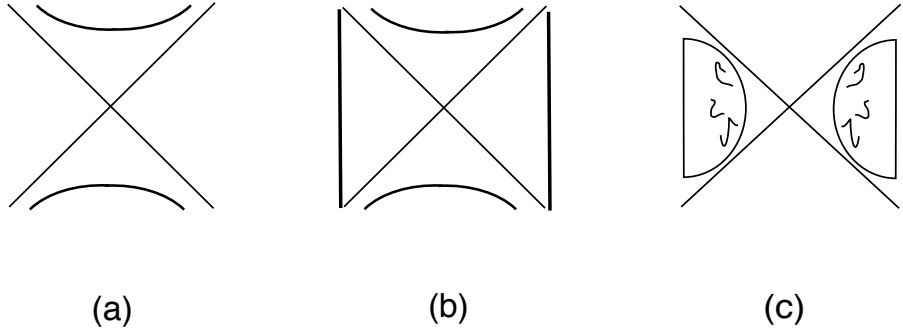}
\caption{(a) The eternal hole, with two asymptotic infinities. (b) The boundary CFT consists of two entangled copies. (c) The state of each copy can be replaced by a gravitational solution, giving a pair of entangled solutions.}
\label{fmal}
\end{center}
\end{figure}

 \begin{figure}[htbp]
\begin{center}
\includegraphics[scale=.45]{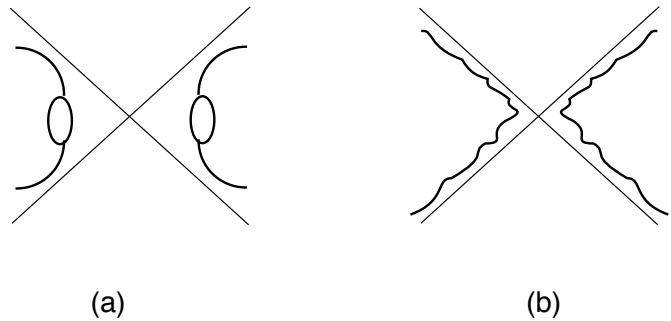}
\caption{(a) If most states were black holes with horizon, then it is unclear what we mean by entangling gravitational solutions. (b) In the fuzzball picture all states end before a horizon forms, so we {\it can} make sense of entangling gravitational solutions.}
\label{fmal2}
\end{center}
\end{figure}

The fully extended Schwarzschild geometry is depicted in fig.\ref{fmal}(a).
This geometry has two asymptotic infinities, rather than the single infinity that 
we normally have in our world. Israel \cite{israel} postulated that we should think of the two sides of this Penrose digram as describing two copies of our gravitational physics, in the same way that we take two copies of a field theory when using the `real time formalism' to study finite temperature  dynamics. Maldacena \cite{eternal} studied the dual CFT description of the eternal hole in AdS space.  This time there are two asymptotically AdS boundaries, and we should associate a CFT with each boundary. Using Israel's connection to the real time formalism, Maldacena arrived at the conclusion that the CFT state describing the eternal hole  is an {\it entangled} state  of the form
\be
|\psi\rangle=\sum_k e^{-{E_k\over 2T}}|E_k\rangle_L\otimes |E_k\rangle_R
\ee
where the two copies of the CFT from the two sides of the hole are seen to be entangled. These two entangled boundaries are depicted in fig.\ref{fmal}(b).

Van Raamsdonk has recently taken this notion of entanglement further, to the entanglement of {\it gravitational} solutions. Consider one of the two copies of the CFT in Maldacena's description. Each state $|E_k\rangle_L$ should be dual to some gravitational solution $|g_k\rangle_L$, and similarly $|E_k\rangle_R$ should be dual to a gravitational solution $|g_k\rangle_R$. Thus we should be able to write the eternal black hole geometry as an entangled sum of {\it gravitational} solutions 
\be
|g\rangle_{eternal}=\sum_k e^{-{E_k\over 2T}}|g_k\rangle_L\otimes |g_k\rangle_R
\label{qwfourt}
\ee
This is interesting, since the spacetime on the  LHS is a  geometry that is {\it connected}  between its left and right sides, while the gravitational solutions appearing on the RHS have {\it no} connection between the L and R sets. Thus we conclude that if we take disconnected gravitational manifolds, but entangle their states, then we generate a connection between the manifolds. This replacement of boundary states by their corresponding gravitational solutions is depicted in fig.\ref{fmal}(c).

Returning to the postulate (\ref{qwfourt}), we do notice a potential difficulty. In a theory of gravity, most of the states at an energy $E$ are expected to be {\it black holes}. What geometry $g$ should we take for such states? If we take the traditional metric with horizon (fig.\ref{fmal2}(a)), then this metric can be continued past the horizon, and into another asymptotically AdS region; thus our state $|g_k\rangle_L$ which was supposed to describe a metric with {\it one boundary} now seems to  describe a metric with {\it two} asymptotic  boundaries. But this is where our understanding of black hole microstates helps us.  These microstates are fuzzballs, with no horizon, and so there is no distinction in principle between hole and states that describe `quanta in AdS' (fig.\ref{fmal2}(b)).  Thus the sum (\ref{qwfourt}) does make sense.

\begin{figure}[htbp]
\begin{center}
 \includegraphics[scale=.65]{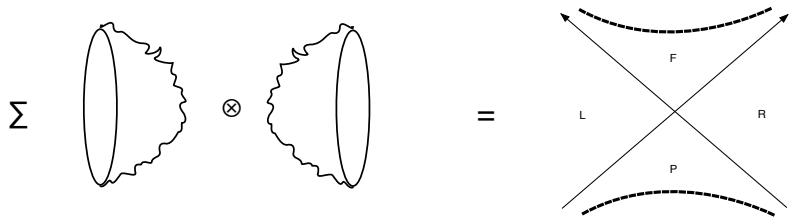}
\caption{{Assuming that all black hole states are fuzzballs, we find that gravitational solutions all end without horizon. Summing these states (fig (a)) should give the  geometry of the extended Schwarzschild hole.}}
\label{fn7}
\end{center}
\end{figure}

In short, we obtain the picture depicted in fig.\ref{fn7}: summing over {\it disjoint} gravitational states reproduces a {\it connected} eternal black hole spacetime \cite{raamsdonk}. 

\section{Entropy of Rindler and de Sitter spaces}

There is an alternative route that we can use to arrive at the above notion of entanglement in gravity; we describe this here as it will enable us to proceed further to the dynamics of fuzzballs.

\subsection{Entropy of Rindler space}
\begin{figure}[htbp]
\begin{center}
\includegraphics[scale=.65]{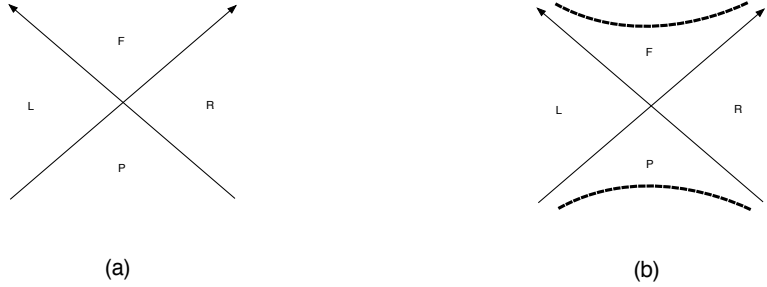}
\caption{{(a) Minkowski space and its Rindler quadrants (Right, Left, Forward and Past). (b) The Penrose diagram of the extended Schwarzschild hole. The region near the intersection of horizons is similar in the two cases.}}
\label{fn5}
\end{center}
\end{figure}

One question that has always puzzled relativists is the following. If black holes have an entropy given by their horizon area, should we associate an entropy with {\it all} horizons? In particular we can take empty Minowski spacetime, and choose `Rindler coordinates' that cover only the `right' quadrant. In these coordinates we see horizons at the boundary of this quadrant (fig.\ref{fn5}(a)). This region of Minkowski space looks very similar to the central region of the full black hole Penrose diagram, fig.\ref{fn5}(b). Should we associate an entropy 
\be
S_{rindler}={A\over 4G}
\label{qone}
\ee
to any area $A$ of  the Rindler horizon? 

For a black hole we could think of the entropy as arising from the number of ways we could make the hole. But in the Rindler case, what would such an entropy be counting? 
Even though the answer to this was not clear, the expression (\ref{qone}) was generally accepted as holding for Rindler horizons, and in particular was used to `derive' Einstein's equation from thermodynamics \cite{jacobson}. What we will now see is that in terms of fuzzballs, there is a logical explanation of (\ref{qone}) as a count of states, even though we are describing the Rindler quadrants of {\it empty} Minkowski space \cite{plumberg}.

Consider the Minkowski spacetime shown in  fig.\ref{fn5}(a), and let the MInkowski coordinates be $x, t$.  Consider a free scalar field $\phi$, and let $|0\rangle_M$ be the vacuum state of this scalar field on Minkowski space. 
 Half of our spacelike slice $x=0$ lies in the left Rindler wedge and half in the right. We can write the complete state $|0\rangle_M$ in terms of states in the left and right wedges
  \be
|0\rangle_M=C\sum_i e^{-{E_i\over 2}}|E_i\rangle_L|E_i\rangle_R, ~~~~~~~C=\Big (\sum_i e^{-E_i}\Big )^{-\h}
\label{split}
\ee
Not surprisingly, the state $|0\rangle_M$ is  entangled between the left and right Rindler wedges. The states $|E_i\rangle_L, |E_i\rangle_R$ are energy eigenstates under the `Rindler time' coordinate in each wedge. For the free scalar field, these eigenstates are explicitly known, though the sum in (\ref{split}) needs regularization at $E_i\r\infty$.

We now make a few observations. First, if we had an interacting scalar field, the states $|E_i\rangle_L, |E_i\rangle_R$ would be eigenstates of the {\it interacting} Hamiltonian. Second, an expansion like (\ref{split}) is expected for {\it any} field, and one field that is always present is nature is the graviton field $h_{ij}$. Let us therefore ask the question that will be central for us: what is the analogue of (\ref{split}) for gravitational fluctuations $h_{ij}$? 

Note that the strength of gravitational interactions increases with energy.  The states $E_i$ have large local energy density  in the region close to the Rindler horizons, due to the large redshift near the horizons in the Rindler metric. Thus the states $|E_i\rangle_R$ for the gravitational field are expected  to be states with the following characteristics: (i) they should `live' in only the right Rindler quadrant (ii) they will have high local energy density near the Rindler horizons (iii) they will involve very nonlinear gravitational interactions near the horizons.

\begin{figure}[htbp]
\begin{center}
\hskip -1 in \includegraphics[scale=.60]{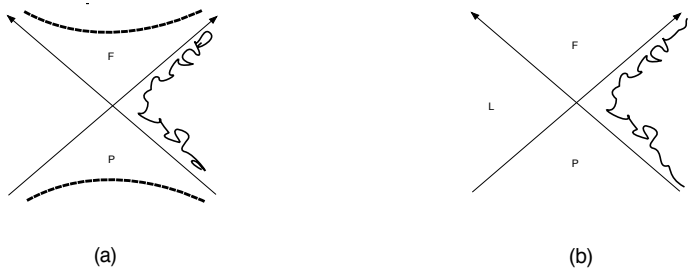}
 \caption{{(a) The eternal black hole spacetime. The geometry of a fuzzball microstate is only the region to the right of the jagged line, and so it lies only in the right quadrant. (b) Taking the limit $M\r \infty$ we obtain fuzzball states lying in the right quadrant of Minlowski spacetime.}}
\label{fn11p}
\end{center}
\end{figure}

But these are just the characteristics of the fuzzball solutions that have been found! The fuzzballs end just outside the the place where the horizon would have occurred. We depict this in fig.\ref{fn11p}(a), where we draw the eternal black hole diagram, and then indicate the boundary of the fuzzball as a jagged line in the right quadrant; thus the fuzzball geometry is only the region to the right of this jagged line. The fuzzball structure in the vicinity of this jagged line  is over very short length scales and very nonperturbative: we have a complicated distribution of  monoples and antimonopoles.  Taking the limit $M\r\infty$ of the black hole mass brings us to Minkowski space, (fig.\ref{fn5}). Thus it is natural to conjecture that the fuzzballs obtained in this limit  are just the solutions $|E_i\rangle_R$ appearing in the decomposition of the Minkowski vacuum $|0\rangle_M$ for the gravitational field. The relation (\ref{split}) for the gravitational field now gives a picture for Rindler space similar to the picture fig.\ref{fn7} for the black hole.

We now have a natural conjecture for the meaning of (\ref{qone}). Consider 3+1 Minkowski spacetime, but as a part of full string theory, so that we have 6 compact directions, and the branes etc.\ that are the allowed sources in string theory. Further, consider this Minkowski spacetime as the limit of the central part of an eternal  black hole diagram in the limit $M\r\infty$. 
There will exist gravitational solutions that approach flat space near infinity, but which end in a monopole-antimonpole structure near the Rindler horizons, such that spacetime `ends' before reaching these horizons.  (We will of course also have  the other branes/fluxes around these monopoles needed to obtain the details of the full fuzzball structure). In the 2-charge extremal case the space of such solutions was quantized and found to yield the correct entropy. Though we cannot yet construct the uncharged solutions needed for the Rindler case, it is natural to conjecture that quantizing the space of solutions will yield (\ref{qone}). 

Thus, roughly speaking,  Rindler entropy counts the number of  manifolds without boundary that fill the right Rindler wedge.

\subsection{Entropy of de Sitter space}

 \begin{figure}[htbt]
\hskip .5 in\includegraphics[scale=.45]{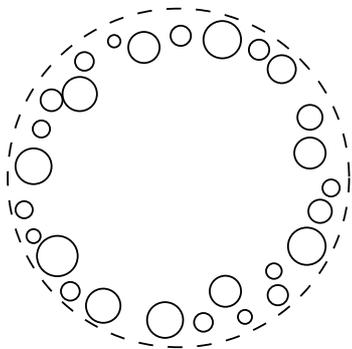}
%
%
\caption{ The horizon of a static patch of de Sitter is indicated by the dashed line. The collection of bubbles indicates a `fuzzball solution' which is de Sitter in its interior but which `ends' in a set of monopole-antimonopole solutions near the horizon. The count of such solutions is conjectured to give the entropy of de Sitter space.  }
\label{fn8}       
\end{figure}

A related question is: how should we understand the entropy of de Sitter space? This spacetime is expanding due to the presence of a positive cosmological constant.  We can choose coordinates in which the metric appears static, and then we get a horizon with an area $A$ at the boundary of this static patch. But this boundary can be moved around depending on our choice of static coordinates, so it was  unclear if we should attach any  meaning  to the entropy
\be
S_{de-Sitter}={A\over 4G}
\label{qtwo}
\ee
But we can now understand this entropy in just the way we understood  Rindler entropy. The complete state straddling both sides of the horizon can be written as an entangled  sum of states on each side. The entropy of de Sitter gives the count of all gravitational solutions that approach regular de Sitter at the center of the static patch, but that end without boundary in a structure of monopoles and antimonopoles before reaching the boundary of the static patch. Thus the entropy (\ref{qtwo})  counts compact manifolds without boundary for the situation where we have a positive cosmological constant (fig.\ref{fn8}).

\section{Summary of comparison between the two notions of holography}

We have seen that the term `holography' has been used in two different contexts in the study of gravity. In gauge/gravity duality, holography maps a gravity theory to an abstract theory at an arbitrarily placed boundary. There are no gravitational degrees of freedom at this boundary itself, its area does not give an entropy, and there is no Hawking radiation from this boundary creating entangled pairs and leading to an information problem. By contrast, the origins of holography lie in  black hole thermodynamics, where the holographic boundary is the horizon. The horizon location is fixed, its area gives the entropy, and radiation from this horizon leads to the information paradox. If we confuse these two notions of holography, then we arrive at an erroneous conclusion that the information paradox is somehow magically evaded if we assume an imaginary surface at the horizon that carries the degrees of freedom of the hole. In actuality, the information paradox is evaded only because there are {\it real} degrees of freedom (hair) at the location of the horizon, and the construction of this hair is accomplished in string theory by the fuzzball construction.

\section{Approximate complementarity for expectation values}

We have seen that in string theory black holes are fuzzballs; i.e. the spacetime ends in a stringy mess just outside $r=2M$, rather than continuing smoothly past the horizon.   We have also noted that for the simple nonextremal microstates that have been constructed, the radiation from ergoregions in the fuzzball exactly matches onto the Hawking radiation expected from those microstates. For typical microstates, these radiated quanta would have energy $E\sim kT$, and we thus see that the Hawking radiation emerges by a process that sees the details of the microstate. Since the radiation is not produced by pair creation at an `information free horizon', we do not have the Hawking paradox that results from ever growing entanglement between the inside and outside of the hole.

Thus with the fuzzball picture we resolve the information paradox. But we can still ask a subsidiary question: is there any meaning at all to the geometry that we traditionally attribute to the  black hole, where we smoothly continue past a horizon and find an interior region?

We now proceed to some conjectures on what this traditional geometry could mean \cite{plumberg,bb}. We will spilt our physical processes into two categories, in line with what we encounter in normal statistical systems. We can understand these two categories by considering a beaker of water.  First we have microscopic physics, like the Brownian motion of a single water molecule in the beaker. The details of this motion depend on the precise choice of microstate for the water. But we also have macroscopic physics, like the response of a thermometer when placed in the water. The mercury level of the thermometer is roughly the same for all generic states, so we can replace the given state of water by the canonical ensemble over microstates. For the fuzzball, we have a similar split of operators:

\b

(a) Those at energies $\sim kT$. These are the analogues of  Brownian motion physics. Each fuzzball radiates $E\sim kT$ quanta differently, and this is how the information in the fuzzball gets encoded in the Hawking radiation.

(b) Those for energies $E\gg kT$. Here it is possible that there is an approximation where we can replace the precise state of the fuzzball by an ensemble average, still obtaining a good approximation of the physics. We will now see that such an approximation leads us back to the traditional geometry of the hole.

\b

\begin{figure}[htbp]
\begin{center}
\includegraphics[scale=.45]{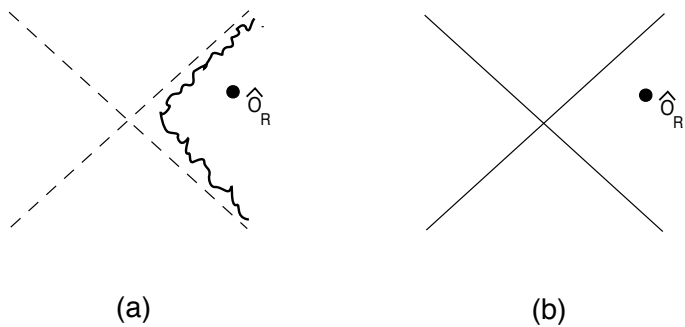}
\caption{{(a) Expectation value of an operator $\hat O_R$  in the right Rindler wedge in a given fuzzball state $|E_k\rangle$. (b) For suitable $\hat O_R$, this expectation value can be approximated by the canonical ensemble over fuzzball states, and thus computed in the full Minkowski space.}}
\label{fn5q}
\end{center}
\end{figure}

 The horizon structure of the eternal hole is similar to the structure of Rindler horizons in Minkowski space (fig.\ref{fn5}), so let us map the black hole problem to the corresponding problem in Rindler space. Observers outside the black hole horizon are described by operators $\hat O_R$ which, in the Minkowski problem, will be located in the right Rindler wedge (fig.\ref{fn5q}(a)). A hole made by collapse is in a definite pure state, which we can take to be the analogue of one of the Rindler states  $|E_k\rangle$. Thus measurements outside the hole correspond to Rindler correlators  ${}_R\langle E_k|\hat O_R|E_k\rangle_R$. There is no evidence of horizon-like behavior so far, since a Rindler state ends in a complicated mess before reaching the location of the horizon; this is depicted in fig.\ref{fn5q}(a). 
 
 The {\it full} Minkowski spacetime does have the Rindler horizons. Suppose we compute the correlator of the same operator $\hat O_R$  as above but in the full Minkowski vacuum; this would be the analogue of computing the correlator in the eternal black hole spacetime with horizons. Noting the decomposition (\ref{split}), we find
\bea
{}_M\langle 0|\hat O_R|0\rangle_M&=&C^2\sum_{i,j}e^{-{E_i\over 2}}e^{-{E_j\over 2}}{}_L\langle E_i|E_j\rangle_L {}_R\langle E_i|\hat O_R|E_j\rangle_R\nn
&=&C^2\sum_i e^{-E_i}{}_R\langle E_i|\hat O_R|E_i\rangle_R
\label{qwe1}
\eea
Thus the expectation value in the Minkowski vacuum is given by  a thermal average over the Rindler states. By contrast, our physical situation needed us to compute the correlator in {\it one} Rindler state: ${}_R\langle E_k|\hat O_R|E_k\rangle_R$.  

  Now we come to a crucial point, which is a basic fact of statistical mechanics:  for a {\it generic} state $|E_k\rangle_R$ and {\it appropriate} operators $\hat O_R$ we should be able to replace expectation values in the state $|E_k\rangle_R$ by an ensemble average
\be
{}_R\langle E_k|\hat O_R|E_k\rangle_R\approx {1\over \sum_l e^{-E_l}}\sum_i e^{-E_i}{}_R\langle E_i|\hat O_R|E_i\rangle_R={}_M\langle 0|\hat O_R|0\rangle_M
\label{qwe2}
\ee
This is depicted in fig.\ref{fn5q}(b). Exactly the same arguments hold for the black hole case since the horizon structure of the extended Schwarzschild hole is the same as that of Minkowski space (fig.\ref{fn5}). 

In short, even though a single fuzzball has no region which is `inside the horizon', we  still find that the expectation value of `appropriate operators'  can be computed to good accuracy by using the {\it traditional} eternal black hole geometry which extends smoothly past a regular horizon. Thus we have obtained an analogue of complementarity,  with two small caveats: (i) it is not a general relation for all operators, but for operators that have an appropriate thermodynamic behavior (ii) it is at its heart an {\it approximate} relation, though the approximation (\ref{qwe2}) would be excellent for large black holes.

\section{Approximate complementarity for dynamical observations}

In the above section we saw how expectation values of appropriate operators in {\it one} fuzzball state could be approximated by expectation values in the ensemble average over fuzzball states, and thus mapped to expectation values in the full extended Schwarzchild geometry. In this section we move towards a more dynamical picture of infall, where we examine the motion of a quntum falling towards the fuzzball surface.

 \begin{figure}[htbp]
\begin{center}
\includegraphics[scale=.95]{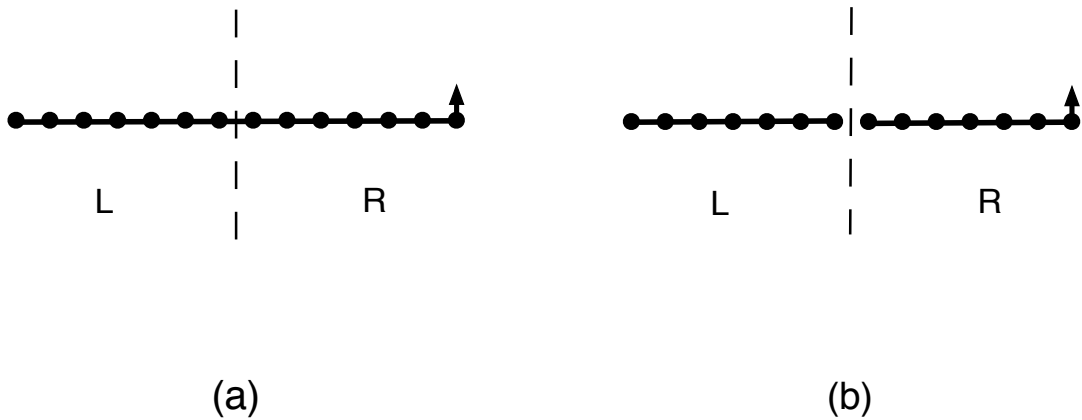}
\caption{{(a) A lattice of spins with a Hamiltonian that makes the spin hop from one site to the next. (b) The same system, but with the link interaction between the L and R sides removed; now a spin cannot hop from one side to the other.}}
\label{fspin2}
\end{center}
\end{figure}

Let us first notice an important difference between entanglement in the black hole case and in the case of normal statistical systems. In fig.\ref{fspin2}(a) we depict a 1-d statistical system of spins. We can place a spin at any lattice site, and we imagine that the Hamiltonian causes spins to hop from any one site to its neighbouring site. This hopping is accomplished, as usual,  by the interaction term between neighbouring sites. We focus on the `left moving' part of the interaction, so the spin in the figure would start hopping to the left.

Let us divide this lattice of spins into a left and a right half; we let the right (R) region cover points $x>0$ and the left (L) region cover points $x<0$.  We can write the ground state of the entire system as an entangled state between the left and right sides
\be
|\Psi\rangle=\sum_i C_i |E_i\rangle_L\otimes |E_i\rangle_R
\label{pone}
\ee
just as we have been doing in the above sections. If we compute an expectation value of an operator $\hat O_R$ in the R sector, then we will find an ensemble average over $|E_i\rangle_R$ states, as before. But now consider a dynamical question. Suppose we start with a spin on the R side, which is hopping leftwards towards $x=0$. If the R and L sectors were totally decoupled, then the spin could never enter the L region. But  for the system in fig.\ref{fspin2}(a) we see that the two sectors are {\it coupled}. The full Hamiltonian has the form
\be
H=H_L+H_R+H_{int}
\ee
where $H_L, H_R$ involve only L and R variables respectively, but $H_{int}$ couples these degrees of freedom. This coupling arises from the link on the lattice that joins the two spins closest to $x=0$ on the L and R sides.

 \begin{figure}[htbp]
\begin{center}
\includegraphics[scale=.95]{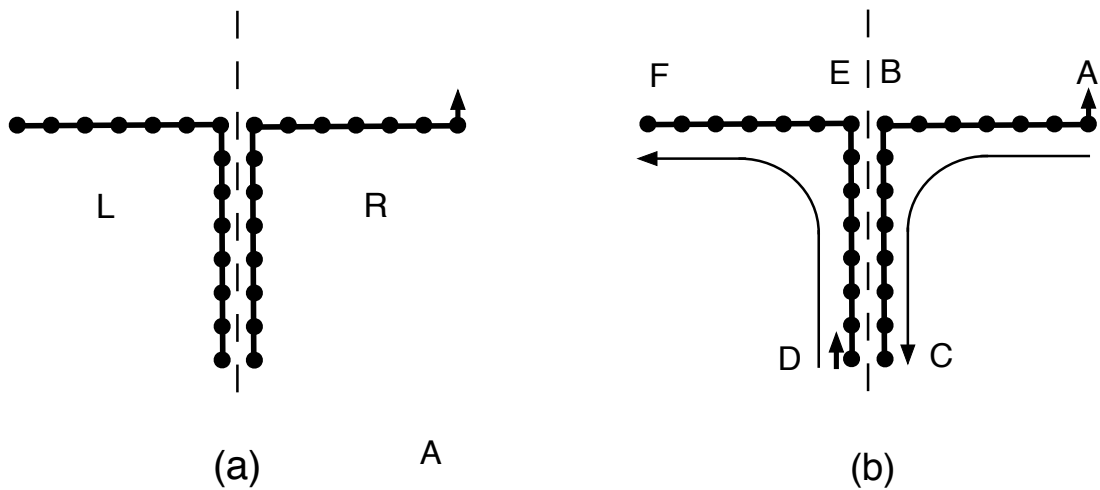}
\caption{{(a) The L and R sides are not connected, but each has a `reservoir' of sites which can trap spins for long times. (b) A spin staring at A gets trapped in the R reservoir, while we arrange for an identical spin to emerge from the L reservoir and continue to the left, mimicking the appearance of continuous motion from the R side to the L side.}}
\label{fspin3}
\end{center}
\end{figure}

But if we return to our gravitational system, we see no such coupling. The states $|E_i\rangle_R$ are fuzzballs that end smoothly; there is no place to `go through' and link up with some other fuzzball in the L sector. But if there is no link to the L sector, then how do we imagine any approximation where we do invoke a spacetime that continues past the horizon?

To see the essential idea that we will propose, let us return to our lattice of spins and cut the link between the L and R sectors (fig.\ref{fspin2}(b)). Now the Hamiltonian has the form
\be
H=H_L+H_R
\ee
and there is indeed no interaction between the L and R sectors. In this situation a spin starting in the R sector and heading towards $x=0$ will bounce back and return to larger $x$ values. In particular there are no degrees of freedom to `hold' the spin near $x=0$ for large times. A black hole, on the other hand, has a very large entropy concentrated in fuzzball states just outside $r=2M$. To mimic these degrees of freedom, we modify our spin model as shown in fig.\ref{fspin3}(a). The L and R sectors are still uncoupled. But a spin coming in towards $x=0$ from the R sector can 
move into a set of degrees of freedom placed near $x=0$. This `reservoir' of states is large but finite. With this situation, the spin coming in from the R region towards $x=0$ does not immediately return back to larger $x$. But it {\it will} eventually return; after it reaches the bottom of the vertically drawn set of spins, it will bounce back up and return to the R region $x>0$. This situation therefore mimics Hawking radiation from the fuzzball; incoming quanta are trapped for large times by the fuzzball degrees of freedom near $r=2M$, but since the number of degrees of freedom is finite, the information in the infalling quanta  is eventually returned to infinity by radiation from the fuzzball surface. 

We imagine a  similar reservoir of states for the L sector, as shown in the figure. Now let us see how with this situation we can mimic the motion of a spin from the R sector to the L sector. The ground state  (\ref{pone}) of a typical system has a very particular relation between the L and R states. When we take a spin excitation in the R sector, we will again take a state that is carefully correlated with the state in the L sector. If the spin starts $n$ units to the right of $x=0$ in the R sector, then in the L sector we place a spin that in $n$ units deep in the  {\it reservoir} set of states (fig.\ref{fspin3}(b)). The spin in the R sector starts at position A, and hops over to position B. It cannot continue further left, but it travels down into the reservoir to position C. In the L sector, we had placed a spin in the reservoir at position D. This spin moves up to position E, and out left to position F. 

Note that the position of the spin at D is chosen such that it appears at E at just the correct time to mimic a spin that would have hopped over from position B. Thus even though the R and L sectors are disconnected, we have managed to create a state which mimics hopping of a spin from the R to the L sector.  The point of course is that it is not the same spin that moved from the R side to the L side, The spin on the R side was caught by the reservoir on the R side, and the reservoir on the L side  spit out a spin at just the right moment so that it seemed that the spin continued its motion past $x=0$. 

 \begin{figure}[htbp]
\begin{center}
\includegraphics[scale=.85]{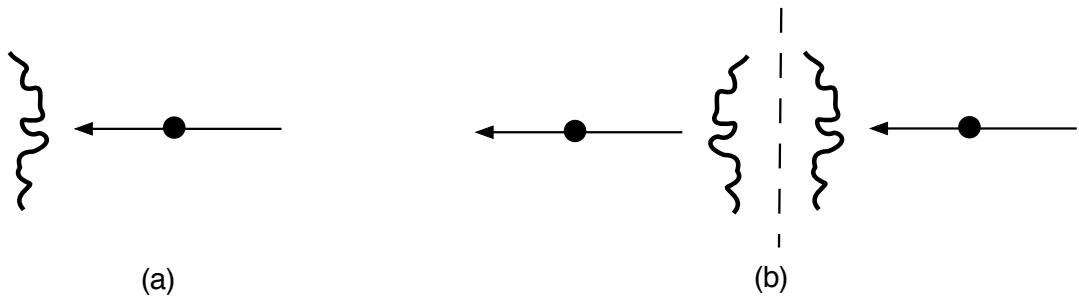}
\caption{{(a) A particle falls onto the fuzzball surface and gets absorbed in the fuzzball degrees of freedom. (b) We consider an auxiliary fuzzball which emits a similar particle at just the correct time so that the trajectories of the two particles make two segments of the same straight line trajectory is spacetime.}}
\label{ffuzz1}
\end{center}
\end{figure}

Now let ask if we can do the same trick with gravity, where the reservoir of states will be the fuzzball degrees of freedom at the horizon. We have noted that the $M\r \infty$ limit of the eternal black hole gives Minkowski space; let us work with this limit to make the equations below simpler. Then the black hole exterior becomes Rindler space. We choose coordinates $t, x$ for Minkowski space. 

The picture we have is depicted in fig.\ref{ffuzz1}. The boundary of the fuzzball is at a constant position trajectory just outside the horizon. We wish to consider a particle falling onto the fuzzball surface. The particle gets absorbed into the fuzzball degrees of freedom, but we wish to have another fuzzball surface, disconnected from the first, from which we can emit another particle. We wish to choose the location of this second surface (and the time of emission) such that the trajectory in the initial space and trajectory in  the second space make up two segments of a straight line trajectory in Minkowski spacetime. 

 \begin{figure}[htbp]
\begin{center}
\includegraphics[scale=.45]{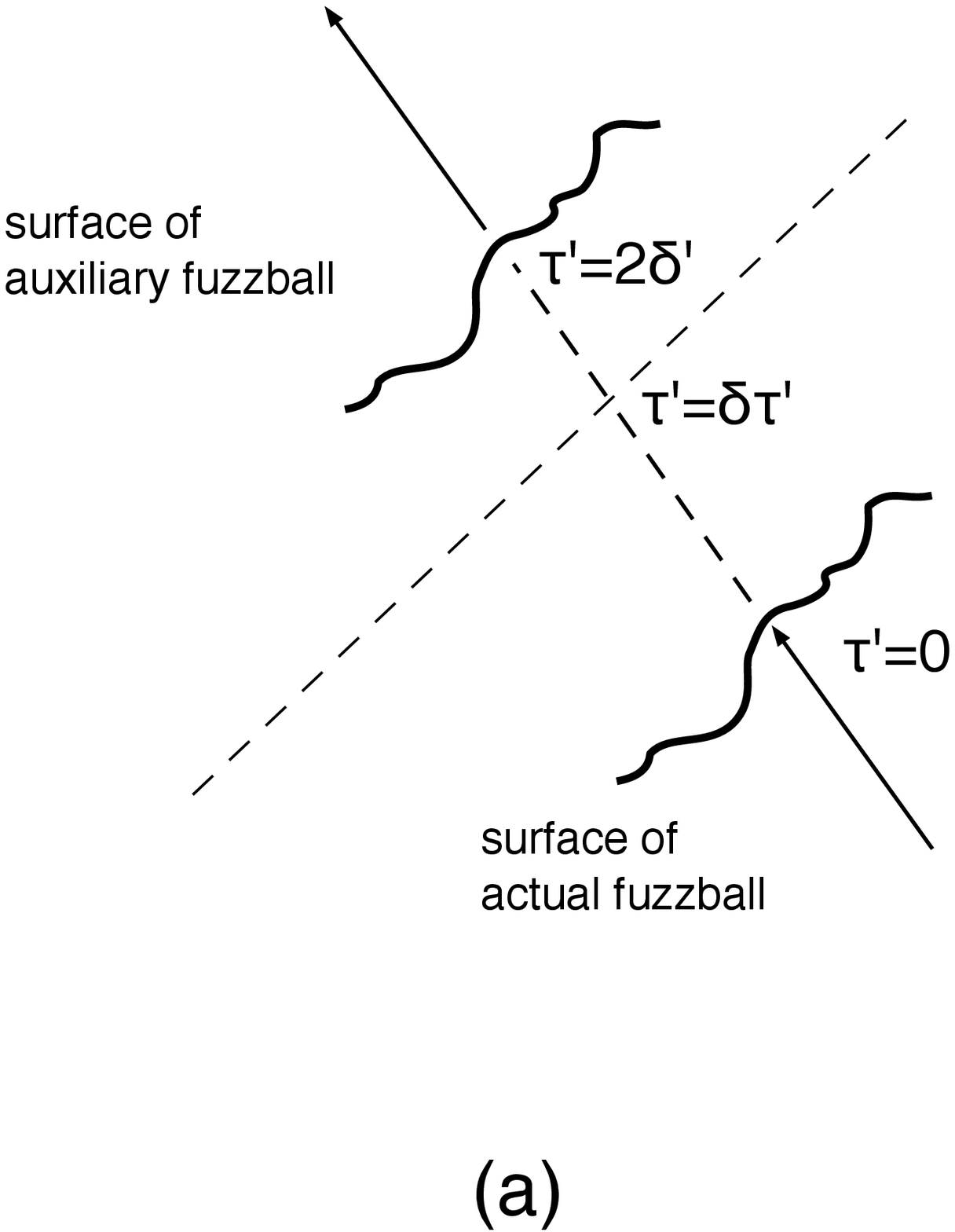}\hspace{3cm}
\includegraphics[scale=.45]{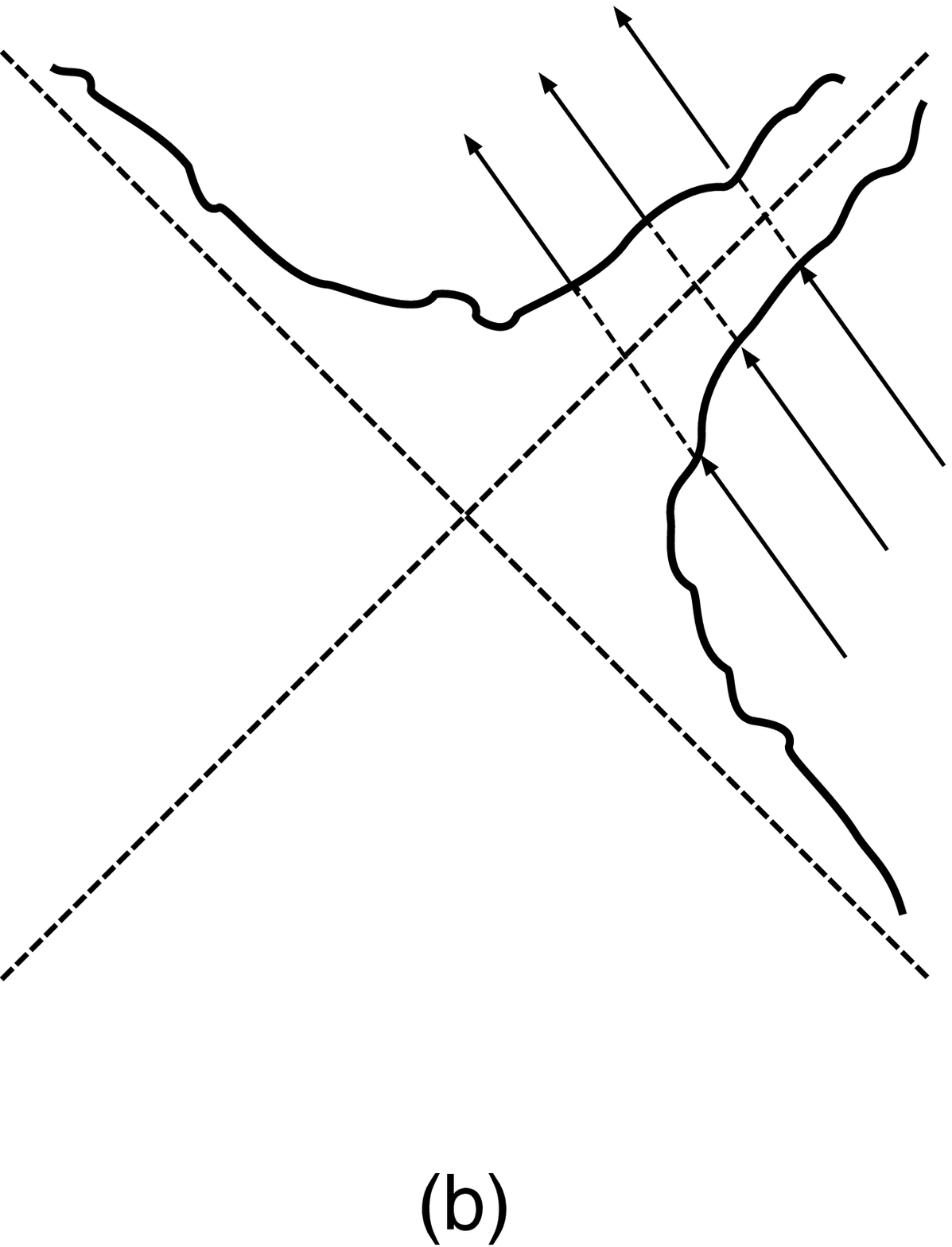}
\caption{{(a) The dashed line depicts the horizon, and the solid curves are the fuzzball surfaces. An infalling quantum is absorbed by the actual fuzzball, but its trajectory is continued later by emission from the auxiliary fuzzball. (b) A family of quanta with the same 4-velocity hit the fuzzball. The spacetime locations for re-emission from the auxiliary fuzzball form a hyperbola in the forward Rindler quadrant.}}
\label{ffuzz3}
\end{center}
\end{figure}

To be more explicit, let us consider the spacetime diagram for such a construction, depicted in fig.\ref{ffuzz3}(a). Let $\tau'$ denote the proper time along the trajectory of the infalling particle. We let $\tau'=0$ at the point where the particle meets the surface of the actual fuzzball. In the embedding Minkowski spacetime, the particle would have continued for a proper time $\delta \tau'$ further before reaching the horizon $t=x$. To make our construction symmetrical between the actual fuzzball and the auxiliary fuzzball, we allow an additional proper time $\delta \tau'$ to elapse along this trajectory. At this spacetime position we imagine placing the second (auxiliary) fuzzball surface. We choose the state of this auxiliary fuzzball such that a particle will be reemitted from it at the appropriate time  and continue along the required particle trajectory in Minkowski space.  
 
 To summarize, we wish to  consider a straight line trajectory in Minkowski space which crosses a Rindler horizon. We break up this motion into two parts: a part in the right Rindler wedge where the particle gets absorbed by a fuzzball surface, and a part in the forward Rindler wedge where a second fuzzball surface reemits a similar particle, creating the illusion of continuous motion across the horizon. In the section below,  we examine in a little more detail the location of the second fuzzball surface and the region of spacetime it creates.

\section{The shape of the recreated region}

In this section we will argue that the surface of the auxiliary fuzzball that we use to recreate the motion past the horizon is a hyperbola in the forward Rindler quadrant (fig.\ref{ffuzz3}(b)). To do this, we will consider a family of infalling particles, all hitting the fuzzball surface with the same relative velocity but at different times. For each such infall trajectory, we will locate the spacetime point $t, x$ in Minkowski space where we must place the auxiliary fuzzball surface in order to recreate the continuation of the infalling  trajectory. The set of points $t, x$ thus found (one for each infalling trajectory) will form a hyperbola in the forward Rindler wedge. Thus the recreated evolution will be in this forward Rindler wedge. 

In Minkowski coordinates $t,x$ the actual fuzzball surface is given  by a curve in the right Rindler wedge
\be
t=\epsilon\sinh {\tau\over \epsilon}, ~~~x=\epsilon\cosh{\tau\over \epsilon}
\ee
Here $\epsilon$ is a small parameter giving the distance of the fuzzball surface from the horizon, and $\tau$ marks proper time along the fuzzball surface. The 4-velocity of a point on the fuzzball surface is (we suppress variables transverse to $t, x$)
\be
U^t={dt\over d\tau}=\cosh{\tau\over \epsilon}, ~~~U^x={dx\over d\tau}=\sinh{\tau\over \epsilon}, ~~~-(U^t)^2+(U^x)^2=-1
\ee
Now we wish to consider a particle falling onto this fuzzball surface. This particle will be travelling in a straight line in Minkowski space; in the corresponding black hole problem this would be a geodesic naturally described in Kruskal coordinates. In Minkowski space the trajectory of the infalling particle has 4-velocity
\be
V^t=\alpha, ~~~V^x=\beta, ~~~\alpha^2-\beta^2=1
\label{ptwo}
\ee
We write
\be
\alpha=\cosh\nu, ~~\beta=\sinh\nu
\ee
where $\nu$ is a constant for the trajectory. We have $\nu<0$ since the particle is coming in from the right.  Let the proper time along the fuzzball surface at the point of impact be $\tau=\tau_c$. The strength of the impact is described by
\be
(U^a-V^a)^2=4\sinh^2{{\tau_c\over \epsilon}-\nu\over 2}
\ee
We write
\be
{\tau_c\over \epsilon}-\nu\equiv F
\ee
and hold $F$ fixed for all infalling trajectories; thus they will all hit the fuzzball surface with the same impact. Note that we impact the fuzzball surface from the right; this corresponds to  $F>0$. 

Let $\tau'$ be the proper time along the infalling particle trajectory (\ref{ptwo}). We chose the origin of $\tau'$ so that $\tau'=0$ at the moment of collision with the fuzzball surface. Then the wordline of the infalling particle is 
\be
t=\cosh\nu ~\tau'+t_c, ~~~x=\sinh\nu~ \tau' +x_c
\label{pthree}
\ee
where
\be
t_c=\epsilon\sinh{\tau_c\over \epsilon},~~~x_c=\epsilon\cosh{\tau_c\over \epsilon}
\ee
are the Minkowski coordinates of the point of collision with the fuzzball surface.
 
 The horizon is at $t=x$. The value of $\tau'$ when the particle trajectory crosses the horizon is given by setting $t=x$ in (\ref{pthree}); this gives
 \be
 \tau'=\epsilon e^{-({\tau_c\over \epsilon}-\nu)}=\epsilon e^{-F}\equiv \delta \tau'
\ee
This quantity $\delta \tau'$ gives the proper time that would have elapsed on the particle trajectory if it were to move from the point of collision outside the horizon to the horizon itself. We mark off another segment of proper length $\delta \tau'$ on the trajectory so that we are now `equally far' from the horizon on the other side of the horizon. This later point had the value
\be
\tau'=2\delta \tau'=2\epsilon e^{-F}
\ee
for the proper time along the worldline. Using (\ref{pthree}) we find the Minkowski coordinates $(t_f, x_f)$ of this final point
\be
t_f=2\epsilon e^{-F}\cosh\nu +t_c, ~~~x_f=2\epsilon e^{-F}\sinh\nu+x_c
\ee
or more explicitly
\be
t_f=2\epsilon e^{-F}\cosh({\tau_c\over \epsilon}-F) +\epsilon\sinh{\tau_c\over \epsilon}, ~~~x_f=2\epsilon e^{-F}\sinh({\tau_c\over \epsilon}-F)+\epsilon\cosh{\tau_c\over \epsilon}
\ee
Finally we wish to examine the set of points $(t_f, x_f)$ that are obtained for different choices of the collision point $(t_c, x_c)$ on the fuzzball surface. We find the relation
\be
t_f^2-x_f^2=\epsilon^2[1+2 e^{-2F}]
\ee
so the points that recreate the other side of the horizon lie along a spacelike hyperbola, and the quanta emitted from these points cover the forward Rindler wedge (fig.\ref{ffuzz3}(b)). 

\section{Limits to spacetime re-creation}

We have placed a set of auxiliary fuzzballs in a hyperbola in the forward wedge of Rindler space (fig.\ref{ffuzz3}(b)). The particle trajectories emerging from these auxiliary fuzzballs travel in this forward wedge. This recreates motion in the forward wedge, but at this point we should ask: can we create such motion for an arbitrarily long time?

Consider the statistical model of fig.\ref{fspin3}(b). We have mimicked the motion past $x=0$ by putting an appropriate set of spins in the reservoir on the left (L) side. This reservoir is large, but finite. If it was 10 lattice sites deep, for instance, then we can set initial conditions so that we get any desired evolution for upto 10 time steps of hopping. But we cannot mimic the evolution for longer than that.

Returning to the gravity problem, we see that we  have two different situations:

\b

(i) The case where the entropy of the fuzzball is infinite. This is the case for Rindler space, or for the hyperbolic black hole \cite{emparan, raamsdonk2}. In this case we have an infinite number of degrees of freedom, and can reconstruct the trajectories in the forward wedge with arbitrary accuracy. We can therefore expect that we can follow these trajectories for infinite time.

(ii) The case where the entropy of the fuzzball is finite. This is the case for the black hole, or for de Sitter space. In this case we can reconstruct the trajectory only to some approximation, and therefore expect that we can only follow these trajectories for a finite time. Interestingly, in these cases we note that the spacetime geometry of the forward wedge terminates in a singularity after some time, with this time increasing with the number of degrees of freedom present in the fuzzball (Bekenstein entropy). 

\b

These observations suggest a fundamental origin of the usual singularity theorems of general relativity. Once we make a closed trapped surface, the geodesics inside the surface must terminate in a singularity after a finite proper time. We now see that in the picture of complementarity that we have developed with auxiliary fuzzballs, the occurrence of such singularities may be a direct consequence of the finiteness of the number of degrees of freedom present in the fuzzball.

\section{Complementarity and cloning}

One question that arises immediately in any discussion of complementarity is the following. If we can emit the information of an infalling quantum to infinity, and also have a copy of it fall past the horizon, then have not performed quantum `cloning' which is impossible on grounds of quantum mechanical linearity?

To see how this issue gets bypassed in our construction, consider the statistical model of fig.\ref{fspin3}(b). The spin at A travels down to point C. But another copy of the spin placed at D emerges and travels to F. There is of course nothing wrong with duplication in this fashion: we have {\it started} with two copies of the spin, one at point A and one at point D. We have not taken one copy of the spin and required it to split into two copies -- that would be cloning and is impossible. 

In short, if we make our auxiliary fuzzball to have a state that is carefully crafted to reflect the physics in the right Rindler wedge, then we are not cloning, and there is no problem. Any time two copies of a space are entangled, as in (\ref{split}), there is a precise correlation between the states of the two copies, but this {\it not} cloning. All we have done is extended such a map to states excited above the vacuum, and thus obtained a dynamics that continues past the horizon in the manner depicted in fig.\ref{ffuzz3}(b). 

The question that we could still ask is: {\it why} should we try to make such an extension? Should we not just say that the dynamics of the infalling particle ended at the fuzzball surface when it got absorbed into the fuzzball degrees of freedom? For the black hole, we could indeed say this and stop; having modified the dynamics at the horizon by order unity we have resolved the information paradox. But we have used fuzzballs in a second way -- in the context of building spacetime by entanglement following the proposal of van Raamsdonk \cite{raamsdonk}. If we wish to have an entanglement relation like fig.\ref{fn7}, then we need to be sure that all states in gravity are fuzzballs like fig.\ref{fmal2}(b) and it is not the case that some states have {\it horizons} like fig.\ref{fmal2}(a). Assuming that all states are fuzzballs, we find that we can break up spacetime at {\it any} place into a complete set of fuzzballs; this is just like partitioning Minkowski space into its Rindler sectors around an arbitrary location. In this situation we must find a way to reconstruct the full spacetime from just one Rindler wedge, and this is where the continuation of spacetime using the auxiliary fuzzball comes in.

\section{Complementarity and AdS/CFT}

In the early days of fuzzballs a question that was commonly asked was the following. If the spacetime ends at the horizon, then are we saying we go `splat' at the horizon? If so, then how can we reconcile this with any picture where we expect to fall smoothly through the horizon?

Complementarity seeks to achieve both these outcomes, but there was no way to realize complementarity in ordinary gravity. But now that we have seen a way to realize a kind of complementarity through fuzzballs, let us step back and ask more generally how one can go `splat' and yet be `unchanged' in another description. To illustrate the general notion we take a simple example: that of AdS/CFT duality. 

\begin{figure}[htbp]
\begin{center}
\includegraphics[scale=.45]{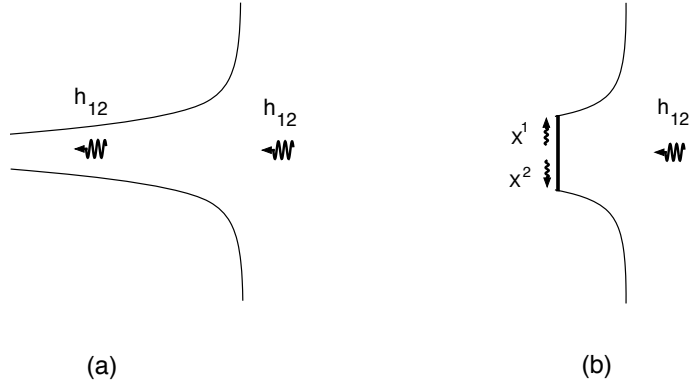}
\caption{(a) The geometry created by D1D5 branes. A graviton $h_{12}$ passes smoothly into the AdS interior. (b) The AdS region is replaced by the D1D5 brane CFT. The graviton gets converted into a pair of vibrations on these branes moving in opposite directions, but all the information of the graviton is preserved.}
\label{fads}
\end{center}
\end{figure}

In fig.\ref{fads}(a) we have the geometry generated by  a bound state of D1 and D5 branes. The geometry at infinity is Minkowski space compactified on $T^4\times S^1$. Near the branes  there is a `neck', and then an AdS region. A graviton with indices in the torus is a scalar in AdS; let us take this graviton to be $h_{12}$ for concreteness where $z_1, \dots z_4$ are the direction on $T^4$.  

In fig.\ref{fads}(b) we depict the picture where we use the open string description of the D-branes; the open strings give vibrations of the branes which form the dual CFT \cite{maldacena}. The graviton $h_{12}$ hits the D1D5 brane state, getting converted to open string excitations on these branes. At leading order in the coupling, $h_{12}$ brakes into a pair of vibrations in the $X^1$ and $X^2$ directions.  One can say that the graviton has gone `splat' to such an extent that it has decomposed into two parts.

But in the description of fig.\ref{fads}(a) the graviton has passed deep into the AdS region, without distortion. How can the graviton go splat, and yet in some sense `feel no change'? 

The reason is not hard to find. What has happened is that the Hilbert space of states of the graviton $h_{12}$ has mapped faithfully into the Hilbert space of vibrations $X^i$ of the D1-D5 brane system \cite{comparing}. If we map vectors from one space to another without changing dot products between vectors, then there is no alteration to the dynamics, though the physical realization of the system may look completely different. The fact that the map is faithful (or close to faithful) is due to the fact that the number of degrees of freedom in the D1D5 system is large, and we get `fermi-golden rule' absorption, which is insensitive to the precise energy levels; it depends only on the density of states and the average coupling. (See \cite{plumpre,plumberg} for a more detailed discussion of absorption.)

We can look more explicitly at how dynamics is encoded in these two descriptions \cite{lm4}. Suppose we throw in a graviton $h_{12}$, and a little while later, a graviton $h_{34}$. The first graviton breaks into $X^1, X^2$ and these open string excitations start to separate from each other (each moving at the speed of light) along the $S^1$ direction shared by the D1 and D5 branes. The second graviton breaks into excitations $X^3, X^4$, and these start to separate in a similar manner. In the gravity description of fig.\ref{fads}(a), we can say that the system consisting of the two gravitons is `undistorted' as long we maintain the separation between the gravitons. In the CFT description we can recover the value of this separation by looking at the separation between $X^1, X^2$ and subtracting the separation between $X^3, X^4$. Thus the two gravitons have gone `splat' on the D1D5 branes, and yet managed to preserve all the structure that corresponds to their being `undisturbed'. 

\begin{figure}[htbp]
\begin{center}
\includegraphics[scale=.45]{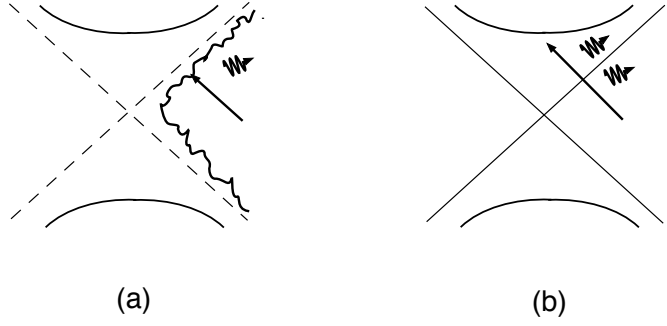}
\caption{(a) Real degrees of freedom at the horzion emit information in Hawking radiation. An infalling quantum is caught by the fuzzball and one may construct an auxiliary spacetime to continue iys motion inside. (b) We cannot solve the information puzzle if we take the literal analog of AdS/CFT. This time we have no real degrees of freedom at the horizon, but try to make a dual description at the Rindler boundary.}
\label{fnf}
\end{center}
\end{figure}

Let us see what we can and cannot borrow from this AdS/CFT example for our fuzzball dynamics. The idea that we can go `splat' in one description, and yet continue smoothly in another description, is similar in the two cases. The difference is that in the fuzzball case we have real degrees of freedom at the fuzball surface that encode the data of the fuzzball. This is depicted in fig.\ref{fnf}(a), where the details of the jagged line carry this data. The emerging radiation thus carries the information of the fuzzball. When high energy quanta hit the fuzzball (shown by the ingoing arrow in fig.\ref{fnf}(a)), we can absorb their data in the collective modes of the fuzzball in a manner similar to that in the AdS/CFT duality example, and then construct an auxiliary spacetime as described above to continue the motion to an `interior' region.   But it is important to note that we cannot just take an analog of AdS/CFT and solve the information problem. That would be analogous to fig.\ref{fnf}(b), where we have no real degrees of freedom at the horizon, and try to capture the physics behind the Rindler horizon with a dual CFT placed at the Rindler horizon. 
In that case the information does not come out in the Hawking radiation.

\section{Gravity-gravity duality}

\begin{figure}[htbp]
\begin{center}
\includegraphics[scale=.45]{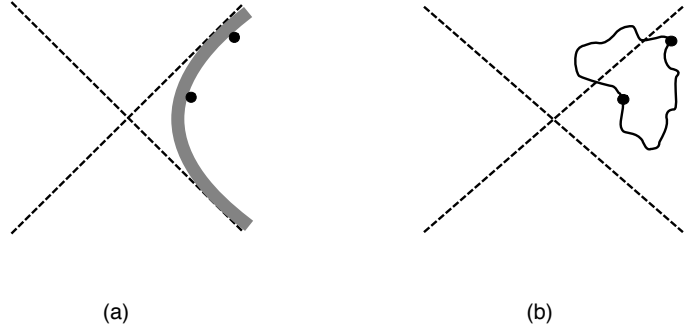}
\caption{(a) Computing a correlator in Minkowski space using Rindler coordinates; we use only degrees of freedom in the right Rindler wedge, and in particular get a contribution from degrees of freedom close to the Rindler horizon. (b) Computing the correlator using Minkowski variables;   the path integral explores both sides of the horizon. }
\label{fads2}
\end{center}
\end{figure}

\begin{figure}[htbp]
\begin{center}
\includegraphics[scale=.45]{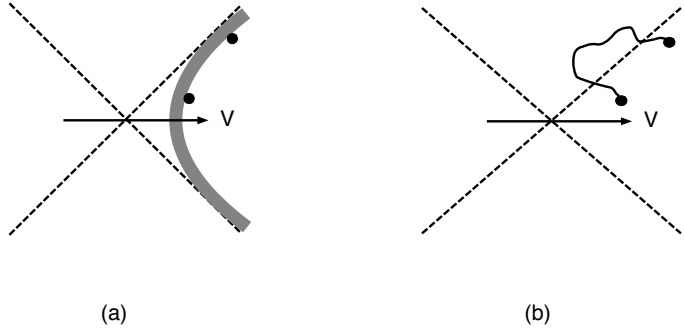}
\caption{(a) Adding  a potential forces the Rindler computation to use only degrees of freedom near the horizon. (b) The Minkowski computation path integral explores only the left side of the diagram.}
\label{fads3}
\end{center}
\end{figure}

Consider Minkowski space divided into its four Rindler quadrants. Consider a free scalar field $\phi$ on this space, and let us see how we would compute the 2-point function
\be
G(x, x')={}_M\langle 0 | T[\hat \phi(x)\hat \phi(x')]|0\rangle_M
\ee
where both points $x, x'$ are in the right Rindler wedge. We further assume that $x, x'$ are both very close to the Rindler horizon, but not close to each other (fig.\ref{fads2}). 

We can compute $G(x, x')$ in two ways. First, we can do the computation using only the states of the right Rindler wedge. Writing
  \be
|0\rangle_M=C\sum_i e^{-{E_i\over 2}}|E_i\rangle_L|E_i\rangle_R, ~~~~~~~C=\Big (\sum_i e^{-E_i}\Big )^{-\h}
\label{splitw}
\ee
 we have
 \bea
{}_M\langle 0|T[\hat \phi(x)\hat \phi(x')]|0\rangle_M&=&C^2\sum_{i,j}e^{-{E_i\over 2}}e^{-{E_j\over 2}}{}_L\langle E_i|E_j\rangle_L {}_R\langle E_i|T[\hat \phi(x)\hat \phi(x')]|E_j\rangle_R\nn
&=&C^2\sum_i e^{-E_i}{}_R\langle E_i|T[\hat \phi(x)\hat \phi(x')]|E_i\rangle_R
\label{qwe1w}
\eea
To evaluate the above correlator, we can expand $\hat \phi$ in Rindler modes
\be
\hat\phi(t_R, x_R)=\sum_\omega [f_\omega(x_R)e^{-i\omega t_R}\hat A_\omega + f^*_\omega(x_R)e^{i\omega t_R}\hat A_\omega^\dagger]
\ee
where $t_R, x_R$ are Rindler coordinates. Since the points $x, x'$ have been taken to be well separated, the contribution to the correlator is from low energy modes. 
But since the points $x. x'$ are close to the Rindler horizon, the Rindler states $|E_i\rangle_R$ will be highly populated with these low energy quanta. The expectation value of the number operator $n_\omega$ for quanta with energy $\omega$ in a typical state $E_i\rangle_R$ is 
\be
{}_R\langle E_i|\hat n_\omega|E_i\rangle_R\approx {1\over e^{2\pi\omega}-1}\approx {1\over 2\pi \omega}
\ee
(the last approximation is for energies low compared to the Rindler temperature). We can thus compute the correlator $G(x, x')$ using states that are at high temperature.

But there is a second way to compute the same correlator. We can expand $\phi$ in Minkowski coordinates
\be
\hat\phi=\sum_k [{1\over \sqrt{2\omega_k}}e^{ikx-i\omega t}\hat a_k + 
{1\over \sqrt{2\omega_k}}e^{-ikx+i\omega t}\hat a_k^\dagger]
\label{aone}
\ee
Now the computation is much simpler; we just insert (\ref{aone}) in ${}_M\langle 0|T[\hat \phi(x)\hat \phi(x')]|0\rangle_M$ and obtain the result. This time we see no high temperature states; instead we see just vacuum states.

All this was of course elementary; we can compute in Rindler coordinates or in Minkowski coordinates, and will get the same result each time. The point we wish to note is the following. In the Rindler computation, we use a large number of degrees of freedom concentrated near the Rindler horizon; this is indicated by the grey band in fig.\ref{fads2}(a). The Minkowski computation is not confined to such a band; if we think in terms of path integrals, the paths summed over cross over the Rindler horizon as shown in fig.\ref{fads2}(b). 

Now think of the same situation but with the gravitational field in place of the scalar field. The Rindler states are just the fuzzball states discussed above.  We are computing the correlator of two operators near the horizon. In one computation we use many degrees of freedom -- the fuzzball states -- near the horizon, and in the other we use the whole space but with no degrees of freedom visible near the horizon. The situation resembles what we see in AdS/CFT duality. The CFT computation involves many degrees of freedom at the boundary. The  gravity computation uses the interior of this boundary but not any degrees of freedom at the boundary itself. What is different from usual AdS/CFT is that  both the descriptions we are comparing are in {\it gravity} itself: we can either use the full set of fuzzball degrees of freedom near the horizon surface, or the smooth spacetime past this surface. Thus we have a gravity-gravity duality.  

There is one fact that we have not been careful about above. Consider the Rindler computation. Because we took the points $x, x'$ to be well separated, we got a contribution from near the horizon, but we will also in general get a contribution from paths that wander away from the horizon into the right Rindler wedge. Thus we cannot really say that the Rindler computation is confined to a surface near the horizon. To force the particle paths to stay confined near the horizon, let us imagine a potential that pushes quanta to the left (i.e. negative $x$). This is depicted  in fig.\ref{fads3}(a). Now we will indeed have a computation that is confined to a lower dimension surface in spacetime. The computation using Minkowski variables also involves particle paths that do not venture far into the region $x>0$; this is depicted in fig.\ref{fads3}(b). Now we do have a close analog of AdS/CFT duality, but with both sides of the duality in gravitational variables. We can now ask if the fuzzball degrees of freedom can be interpreted as an effective CFT; in that case we would indeed have obtained an AdS/CFT type map from gravity alone. The natural candidate for the scale factor in such a CFT would be the distance from the Rindler horizon;  we have taken our points $x, x'$ close to this horizon, but the precise distance from the horizon determines the Rindler temperature we see in the Rindler computation, and the value of this local temperature resembles the local scale factor. The potential of the kind indicated in fig.\ref{fads3} arises in particular if we have AdS space; the gravitational field draws quanta to smaller $r$ values, which we take to be towards the direction $x<0$ in the figure.

If we do not have the kind of potential indicated in fig.\ref{fads3}, then we have a somewhat different situation with our `flat space holography'. We can compute correlators in the analog of Rindler variables, or in the analog of Minkowski variables, and the equality of these two is our `duality'. But the Rindler computation (using fuzzballs) will typically use the whole right Rindler wedge, and the Minkowski computation will use the full Minkowski region.

\section{Overview}

\begin{figure}[htbp]
\begin{center}
\includegraphics[scale=.45]{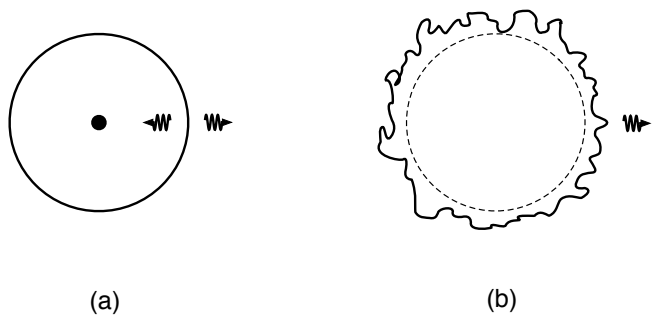}
\caption{(a) The traditional black hole; small corrections at the horizon {\it cannot} get information out in the Hawking radiation. (b) The fuzzball picture of black hole microstates; spacetime ends in stringy theory sources before the horizon is reached. }
\label{fdiss}
\end{center}
\end{figure}

Our story on holography has taken many twists and turns, so it might be helpful to summarize it here:

\b

(a) We noted that the idea of holography that originated from black holes and the idea that is embodied in gauge-gravity duality are somewhat different. In black holes the boundary surface has a fixed location ($r=2M$) and the surface area gives the entropy (\ref{one}). In AdS/CFT the boundary surface can be moved to any location in the AdS region, and the area of this boundary does not give the entropy contained within. 

\b

(b) The traditional black hole has a horizon where physics is `normal' in a `good slicing' (fig.\ref{fdiss}(a)). The Hawking computation showed that this situation leads to information loss/remnants, since pairs produced at the horizon lead to an ever increasing entanglement between the inside and the outside of the hole. It was believed by many people that small corrections to Hawking's leading order computation could remove this entanglement, so we may still be able to preserve the notion of a regular horizon while getting information out in Hawking radiation. But in \cite{cern} it was shown, using subadditivity of quantum entropy, that this hope is {\it false}; small corrections (order $O(\epsilon)$) to Hawking's leading order computation can reduce the entanglement only by a small amount:
\be
{\delta S_{ent}\over S_{ent}}<2\epsilon
\label{fourq}
\ee
Thus the horizon {\it cannot} be a normal place for low energy physics if information is to come out in the Hawking radiation. A corollary of this this theorem is that AdS/CFT duality cannot by itself say anything about the information problem; if we have a black hole with traditional horizon
(like the AdS-Schwarzschild hole) then small quantum gravity effects cannot lead to information recovery. The example of matrix models shows that if we define our gravity theory as the dual of a CFT, then in general {\it we will not get black holes}. Thus the consequence of (\ref{fourq}) is to tell us that the only way to resolve the information paradox is to find a source for order unity corrections to low energy physics at the horizon.

\b

(c) People had looked for deformations of black holes in the past, but the metric of the hole had appeared to be stubbornly unique: this failure to find alternative solutions for given quantum numbers was encoded in the statement `black holes have no hair'. At first string theory does not seem to help; if stringy effects are manifested only near the planck scale, then they would not affect Hawking's argument. But in \cite{emission} it was shown the phenomenon of `fractionation' gives bound states in string theory a size that grows with the number of branes in the state and with the string coupling, in such a way that the size of the bound state wavefunctional is always order horizon size. 

Further progress along this direction is achieved by taking specific bound states and constructing their gravitational solution at the coupling where the black hole is expected. We find that a horizon does {\it not} form; instead we get a fuzzball where the gravitational solution ends just outside the place where the horizon would have been, in a set of string theory sources (KK monopoles, fluxes, branes, etc.). Such solutions are termed fuzzballs, and they give a completely different picture (fig,\ref{fdiss}(b)) for the states of the black hole. 

\b

(d) For the information problem, we need to examine the process by which Hawking radiation is emitted from these fuzzballs. Since there is no horizon, we do not find the `pair creation from vacuum' process that led to Hawking's puzzle. For a set of simple nonextremal microstates it was found that emission occurs from the {\it ergoregions} in the geometry, and the rate of this emission is {\it exactly} the rate expected for Hawking emission from these microstates. 

It is crucial that different microstates have differently placed ergoregions, and the emission is thus dependent on the microstate. This situation should be contrasted with other attempts at the information problem which seek to keep the regular horizon of fig.\ref{fdiss}(a). In many of these approaches one attempts to get the information out by invoking the `high temperature degrees of freedom' that are manifested in the Schwarzschild coordinate frame. It is important to note that such a manifestation of high temperature cannot get the information out; it is simply a reflection of the failure of the Schwarzschild coordinates at the horizon. What we find in the fuzzball construction is {\it genuine} degrees of freedom at the horizon. We cannot obtain these genuine degrees of freedom without the full structure of string theory, which allows us to end the manifold outside the horizon by `pinching off' the compact directions into KK monopoles, and supporting the monopoles by branes/fluxes.   

\b

(e) Finally, we may ask if there is any meaning at all to the traditional black hole geometry, given that all microstates end without horizon. We have argued that for the purposes of an infalling observer, we can construct an auxiliary spacetime where the infalling trajectory can be continued past a horizon. The notion of such a  continuation is related to the idea of van Raamsdonk \cite{raamsdonk} of building spacetime by `entanglement'. This construction looks like a version of complementarity, but it is different in a crucial way.  We do not break up of the original black hole spacetime into different regions across a smooth horizon; instead the spacetime ends outside the horizon in real degrees of freedom, and the continuation is only useful for describing infalling observers for a limited time using auxiliary degrees of freedom.

  \section*{Acknowledgements}

This work was supported in part by DOE grant DE-FG02-91ER-40690. I would like to thank the organizers of COSGRAV12 for their hospitality. I also thank all my collaborators over the years who have contributed to the work presented here.

\end{document}